\documentclass[preprint,showpacs,floats,letterpaper,floatfix,,groupedaddress,eqsecnum]{revtex4}
\usepackage{subfigure}

\usepackage{amssymb,amsmath}
\usepackage[dvips]{graphicx}

\usepackage{dcolumn,epsfig}

\begin{document}

\title{Particle dynamics in a symmetrically driven underdamped inhomogeneous
periodic potential system}
\author{D. Kharkongor$^1$, W.L. Reenbohn$^2$ and Mangal C. Mahato$^1$$^,$}
\email{mangal@nehu.ac.in}
\affiliation{$^1$Department of Physics, North-Eastern Hill University,
Shillong-793022, India}
\affiliation{$^2$Department of Physics, National Institute of Technology 
Meghalaya, Shillong-793003, India}

\begin{abstract}

We numerically solve the underdamped Langevin equation to obtain the 
trajectories of a particle in a sinusoidal potential driven by a temporally 
sinusoidal force in a medium with coefficient of friction periodic in space as 
the potential but with a phase difference. With the appropriate choice of
system parameters, like the mean friction coefficient and the period of the
applied field, only two kinds of periodic trajectories are obtained for all 
possible initial conditions at low noise strengths: one with a large amplitude
and a large phase lag with respect to the applied field and the other with a
small amplitude and a small phase lag. Thus, the periodic potential system
is effectively mapped dynamically into a bistable system. Though the 
directional asymmetry, brought about only by the frictional inhomogeneity, is
weak we find both the phenomena of stochastic resonance, with ready explanation
in terms of the two dynamical states of trajectories, and ratchet effect 
simultaneously in the same parameter space. We analyse the results in detail 
attempting to find plausible explanations for each.
 
\end{abstract}

\vspace{0.5cm}
\date{\today}

\pacs{: 05.10.Gg, 05.40.-a, 05.40.Jc, 05.60.Cd}
\maketitle

\section{Introduction}

In this work we explore the possibility whether an inhomogeneous underdamped 
sinusoidal potential\cite{Risken} system can be an appropriate candidate for 
obtaining ratchet 
current\cite{Magnasco,Prost,Svoboda,Reimann,Julicher,Maddox,Rousselet} while 
simultaneously exhibiting stochastic 
resonance\cite{Benzi,Gamma,Well,McN,Fauve,Roy}. In other words, whether in an 
inhomogeneous system one can obtain an optimally improved response to a 
subthreshold external field as the temperature is raised (stochastic resonance)
in addition to exhibiting net asymmetric transport (ratchet effect) in a 
sinusoidal potential without the application of an external bias or with the 
application of an external field that, on its own, on the average, adds up to 
equal impulse in opposing directions. 

Stochastic resonance was shown earlier to occur in sinusoidal 
potentials\cite{Saikia, Wanda,Liu} contrary to its absence having been stressed
previously\cite{Kim}. It follows from the realization that when driven at a 
suitably large frequency, an underdamped particle moving in the sinusoidal 
potential shows two kinds of response trajectories, one with a large amplitude
(and also large phase lag with respect to the drive) and the other with a 
small amplitude (and small phase lag). These solutions are quite robust and 
have the status of dynamical states. The characteristics that fully establish 
the trajectories as dynamical states are discussed in detail in 
Ref.\cite{Saikia}. The motion (trajectory) of the particle can be in either of 
these two states at any given instant of time.
This is especially clearly observable at lower temperatures before the 
occurrence of stochastic resonance. Therefore, dynamically the system can be 
considered as bistable. The transition between these dynamical states, as the 
temperature is raised, can provide an explanation for the occurrence of 
stochastic resonance in periodic potentials. As far as the phenomenon of stochastic 
resonance is concerned, the sinusoidal potential thus recedes into oblivion 
and the problem is recast into one in a bistable system. This is important, 
for it is in the bistable systems that stochastic resonance is commonly 
observed. 

The two dynamical states and the transitions between them refer mostly to 
trajectories about the minimum within a period of the periodic potential.
However, the statement, "stochastic resonance in a periodic potential", is
meaningful only if the motion includes not only inter-(dynamical)state 
transitions present but also movements between various wells (periods) of the 
potential. In other words, the 'bistability' must be maintained whichever
period of the sinusoidal potential the particle happens to occupy as time 
progresses so that the full sinusoidal potential, and not just one period of
it, is found to be accessed. Given the criterion to be satisfied, the 
occurrence of stochastic resonance is discussed only in terms of the two
dynamical states and any reference to the sinusoidal potential is conveniently
kept silent. However, sometimes, the reference to the interwell transitions
has to be alluded to in order to check the credibility of the claim of 
occurrence of stochastic resonance in periodic potentials. In fact, well
before stochastic resonance occurs, numerous interwell transitions take place.
As the temperature becomes much larger than the temperature at which
stochastic resonance occurs the transitions become so frequent that the two 
dynamical states become hardly discernible with separate identity, for the
intervals of existence of the two states become too short (could even be less 
than a period). 

By an inhomogeneous system we mean, in the present work, a system with 
nonuniform (space-dependent) friction coefficient. It has been theoretically 
shown earlier that an overdamped particle experiences a net effective driving 
force when subjected to a drift in a sinusoidal potential and also a 
sinusoidally modulated space-dependent diffusion, with the same period as the 
potential\cite{Buttiker}. However, diffusion coefficient $D$ depends on both,
the temperature $T$ and the friction coefficient $\gamma$: 
$D\sim\frac{T}{\gamma}$. The effect of temperature variation is considered 
dominant as it determines the local stability of 
states\cite{Buttiker,Landauer}. It has been shown analytically that in an
underdamped periodic potential a similarly periodic variation of temperature
but with a phase difference leads to ratchet current\cite{Blanter}. This
remarkable result was also supported numerically in a related 
work\cite{Benjamin}.  The effect of friction coefficient variation is weak and,
unlike temperature variation, a mere similar periodic variation of friction 
coefficient is not expected to yield ratchet current; to yield ratchet 
current, in this case, one needs to apply, in addition, a zero-mean external 
forcing\cite{Statmech}.

The ratchet effect has been studied in nonuniform friction periodic potential 
systems earlier too but mostly in the overdamped limit\cite{overdamped} unlike 
in the present case of underdamped systems. Also, the period of the drive taken 
earlier used to be very large, typically of the order of mean 
first-passage-time across a potential barrier. In these circumstances, 
stochastic resonance in the periodic potential systems was not found to occur
and hence there was no question of exploring whether both the effects occur
simultaneously in the same parameter space.

As stated earlier, in addition to the ratchet effect discussed above, the 
present work endeavours to observe SR simultaneously in the same parameter
space. Though the present work is not experimental, it is not hard to think of
an easy-to-visualize experimental situation illustrating the present case. 
Consider a series of nodes and antinodes, created by a stationary pressure 
wave, representing sites of alternate low and high density of the medium. A 
particle, therefore, while moving along longitudinally, experiences a periodic 
variation of friction. In addition, if the particle is charged and an 
alternating electric potential field, of same wavelength as the pressure wave 
but with a phase difference, is created the particle will experience a 
spatially periodic force field. The application of a temporally sinusoidal 
external electric field completes the illustrative experimental situation to 
describe the problem at hand and allows the following questions to be asked. 
Is it possible to obtain a net (ratchet) current? If a thermal noise is 
added and increased gradually, either through the external field or the system 
itself, will the response of a host of particles to the external field be 
sharper or more smeared? In other words, does the quality of response, as 
conventionally measured by the (output) signal to (input) noise ratio, 
improve as the noise level is gradually increased? If the 
response shows a peak (stochastic resonance) at an intermediate temperature 
(noise strength) does the net particle current also exhibit a maximum, 
simultaneously\cite{Qian}? These are some of the questions that will be 
attempted to be clarified in a model system. 

These questions together have already been discussed earlier, however, in a 
uniform friction medium and subjected to a biharmonic external 
field\cite{Wanda1}, $F(t)=F_0(A\cos(\omega t)+B\cos(2\omega t))$. In the 
present work, the potential is considered sinusoidal $V(x)=-V_0\sin(x)$ and an 
external field of $F(t)=F_0\cos(\omega t)$. The asymmetry in the system is 
introduced via the phase shifted nonuniform friction 
$\gamma (x)=\gamma_0(1-\lambda\sin(x+\theta))$, with $0\leq\lambda\leq 1$. In 
the case of temperature inhomogeneity, similar in form to the frictional 
inhomogeneity considered here, it has been shown 
earlier\cite{Blanter, Benjamin} that the ratchet current $\overline{v}$ 
depends on the phase difference $\theta$ between the potential function and 
the temperature function as $\overline{v}\sim \sin(\theta)$. Thus, 
understandably, the ratchet current is maximum when the asymmetry is maximum, 
namely, for $\theta=\frac{\pi}{2}$ and the same should be true in the present 
case of frictional inhomogeneity as well.

Interestingly, for $\theta=\frac{\pi}{2}$, the inhomogeneous friction yields
particle trajectories (and stochastic resonance) qualitatively similar to the 
uniform friction 
case in response to the same sinusoidal external field (see Sec. III). However,
the inhomogeneous friction, in addition, helps in obtaining a net (ratchet) 
current. Since, in both the cases, the trajectories show same periodicity as 
the external field, the particle spends almost the same time in the small and 
large friction regions which lie, respectively, on either side of the 
potential minimum. The present situation is therefore quite different from the 
adiabatic drive case wherein the particle is likely to spend unequal durations 
in the two friction regions. Hence the reason put forth for the explanation of 
occurrence of ratchet current in the adiabatic limit will not hold in the 
finite frequency drive case\cite{Statmech}. A closer statistical analysis of 
the parts of trajectories in the two separate regions, however, hints at why 
the net (asymmetric) particle current should be expected.

The turning points of the trajectories on either side of potential 
minimum ($x=\frac{\pi}{2}$) show interesting distributions. At very low 
temperatures, ($T<0.005$), the mean distance of turning points, $dx_r$ (in the
high friction side of the potential peak) and $dx_l$ (in the low friction 
side) from $x=\frac{\pi}{2}$ and their standard deviations, $\Delta x_r$ and 
$\Delta x_l$, about the mean values on both sides are same, showing that the 
motion remains symmetrical on both directions, for all 
$\tau=\frac{2\pi}{\omega}$. However, as $T$ is gradually increased the 
differences show up for $\lambda\neq 0$ and $\theta=\frac{\pi}{2}$. For 
$\tau=7.0$ and at low temperatures only large amplitude trajectories appear 
(see Sec. III, Fig. 9). In this case, one obtains $dx_l>dx_r$ and 
$\Delta x_l>\Delta x_r$ upto a reasonably large temperature. And, for 
$\tau=13.0$ and at low temperatures only small amplitude trajectories appear 
in which case one finds $dx_r>dx_l$ and $\Delta x_l\sim\Delta x_r$. However, 
since inter-well transitions and hence ratchet current is dominantly 
determined by the large amplitude trajectories, there is more likelyhood of 
right-to-left net transport than the other way round. Of course, differences 
of $dx_l$ and $dx_r$, etc. are very small (less than one percent) and 
the interwell transtions are thus decided by the rare large amplitude events.

In the next section II, the model system of our study will be described. In 
section III the numerical results based on our investigations will be presented
in detail and in the last section IV our results will be discussed and 
summarised.

\section{The Model}

We consider the motion of an ensemble of under-damped non-interacting Brownian 
particles each of mass $m$ in a periodic potential $V(x)=-V_0\sin(kx)$. The 
medium in which the particle moves is taken to be inhomogeneous in the sense 
that it offers a spatially varying friction with coefficient
\begin{equation}
\gamma(x)=\gamma_0(1-\lambda\sin(kx+\theta))
\end{equation}
that leads/lags the potential by a phase difference $\theta$. This choice of 
friction coefficient breaks the right-left spatial symmetry of the system.
Here, $\lambda$ is the inhomogeneity parameter, with $0\leq\lambda\leq 1$
and hence $\gamma_0(1-\lambda)\leq\gamma(x)\leq\gamma_0(1+\lambda)$.

In addition, the potential is rocked by a sub-threshold periodic time-dependent
forcing $F(t) = F_0\cos(\omega t)$, with $\omega$ = 2$\pi$/$\tau$ as the 
rocking frequency and $\tau$ as the rocking period. The equation of motion of
the particle subjected to a thermal Gaussian white noise $\xi(t)$ at 
temperature $T$ is given by the Langevin equation\cite{Sancho,Pramana},
\begin{equation}
m\frac{d^{2}x}{dt^{2}}=-\gamma(x)\frac{dx}{dt}-\frac{\partial{V(x)}}{\partial
x}+F(t)+\sqrt{\gamma(x) T}\xi(t).
\end{equation}
with
\begin{equation}
<\xi(t)> =0,
<\xi(t)\xi(t^{'})>=2\delta(t-t^{'}).
\end{equation}
Here, and throughout the text, $<..>$ correspond to ensemble averages.

For simplicity and convenience the equation is transformed in to dimensionless
units\cite{Desloge} by setting $m=1$, $V_0=1$, $k=1$, with reduced variables 
denoted again by the same symbols. Thus, the Langevin equation takes the form
\begin{equation}
\frac{d^{2}x}{dt^{2}}=-\gamma(x)\frac{dx}{dt}
-\frac{\partial V(x)}{\partial x} +F(t)+\sqrt{\gamma(x) T}\xi(t),
\end{equation}
where the potential is reduced to $V(x)=-\sin(x)$ and
\begin{equation}
\gamma(x)=\gamma_0(1-\lambda\sin(x+\theta)).
\end{equation}
Equation (2.4) is numerically solved (i.e., integrated using Ito definition) 
to obtain the trajectories $x(t)$ of the
particle for various initial conditions\cite{Nume,SRS,Mannella}. For initial 
positions $x(t=0)$ the period $-\frac{\pi}{2}<x\leq\frac{3\pi}{2}$ is divided 
uniformly into either $n=100$ or $n=200$ parts (and hence $n$ initial 
positions)and the initial velocity $v(t=0)$ is set equal to zero throughout in 
the following.

\section{Numerical results}
For each trajectory, corresponding to one initial position $x(0)$, the work 
done by the field $F(t)$ on the system, or the input energy, is 
calculated as \cite{Sekimoto}:
\begin{equation}
W(0,N\tau)=\int_0^{N\tau}\frac{\partial U(x(t),t)}{\partial t}dt,
\end{equation}
where $N$ is a large integer denoting the number of periods taken to reach the
final point of the trajectory. The effective potential $U$ is given by
\begin{equation}
U(x(t),t)=V(x(t))-x(t)F(t).
\end{equation}
The mean input energy per period, for the particular trajectory, is therefore 
given by
\begin{equation}
\overline{W}=\frac{1}{N}W(0,N\tau).
\end{equation}

For the deterministic case ($T\rightarrow 0$, for example $T=10^{-6}$) and also 
at low temperatures (low noise strengths, typically less than about 10\% of 
the largest potential barrier corresponding to $F(t)$) the trajectories $x(t)$ 
appear to be periodic and similar 
in nature to that of $F(t)$. Since $V(x)$ is not explicitly dependent on time, 
all of the contribution to $W(0,N\tau)$ comes from the second term in $U$, 
Eq. (3.2), because of the phase difference between $F(t)$ and $x(t)$. The 
nonzero phase difference also implies hysteresis $\overline{x(F)}$, whose area 
is a measure of energy dissipation per period by the system. A typical mean
hysteresis loop is shown in Fig.1. Not surprisingly, therefore, it turns out 
that the corresponding mean hysteresis loop area $\overline{A}$ is same as 
$\overline{W}$. The overall mean input energy per period, $<\overline{W}>$, is 
calculated as an (ensemble) average over all the $n$ trajectories 
(corresponding to all the initial positions considered). Again, we find that 
$<\overline{A}>=<\overline{W}>$. Similarly, the mean velocity $\overline{v}$, 
over one trajectory, is calculated as
\begin{equation}
\overline{v}=\frac{1}{N\tau}x(t=N\tau),
\end{equation}
and the overall mean (net) velocity or the ratchet current, $<\overline{v}>$, 
is calculated as the ensemble average over all the $n$ trajectories. Typically,
as discussed below, $N=200000$ for small temperatures and $N=500000$ for
large temperatures. The value of $N$ is chosen to have relatively small error
bars so that the qualitative features of the results (see below) are not 
obscured. 

\begin{figure}[htp]
\centering
\includegraphics[width=14cm,height=8cm]{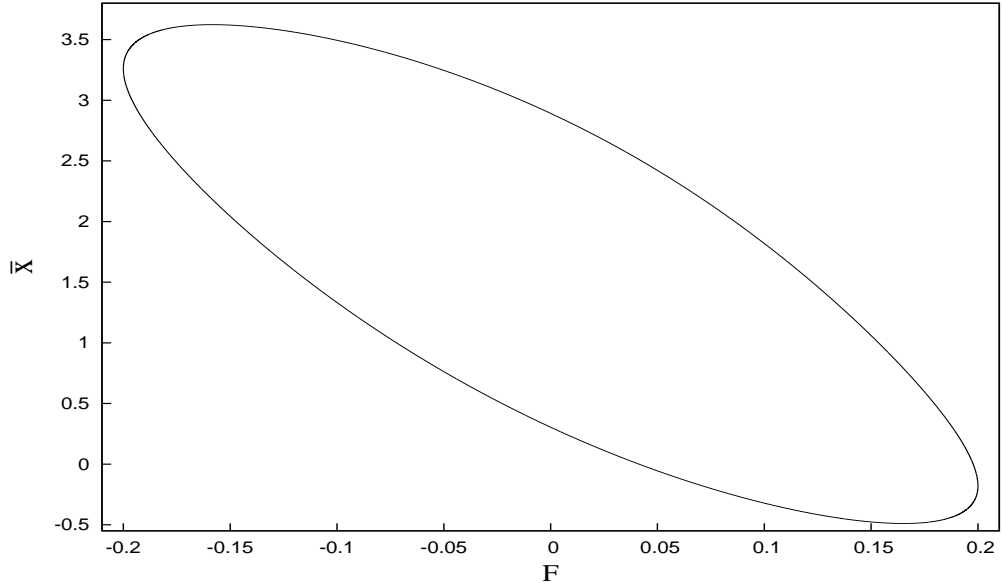}
\caption{Variation of ${\overline{x}}$ as a function of the forcing. Here, 
$\tau$ = 8.0, $\theta$ = $0.5\pi$, $\gamma_0 = 0.07$, $\lambda = 0.9$, 
$T = 0.000001$ and $F_0 = 0.2$. This loop represents a large amplitude state.}
\end{figure}

The error bars are calculated assuming the numerical procedure adopted to be
correct and errors occurring only due inherently to the stochastic dynamics of 
the system. We calculate the $x(N\tau)$ for each initial condition used to 
calculate $\overline{v}=\frac{x(N\tau)}{N\tau}$. The deviations of these 
$\overline{v}$ from the mean ${<\overline{v}>}$ calculated as an ensemble 
average over all trajectories with distinct initial conditions. The standard
deviations $\Delta v$ form the error bars.

\begin{figure}[htp]
\centering
\includegraphics[width=14cm,height=8cm]{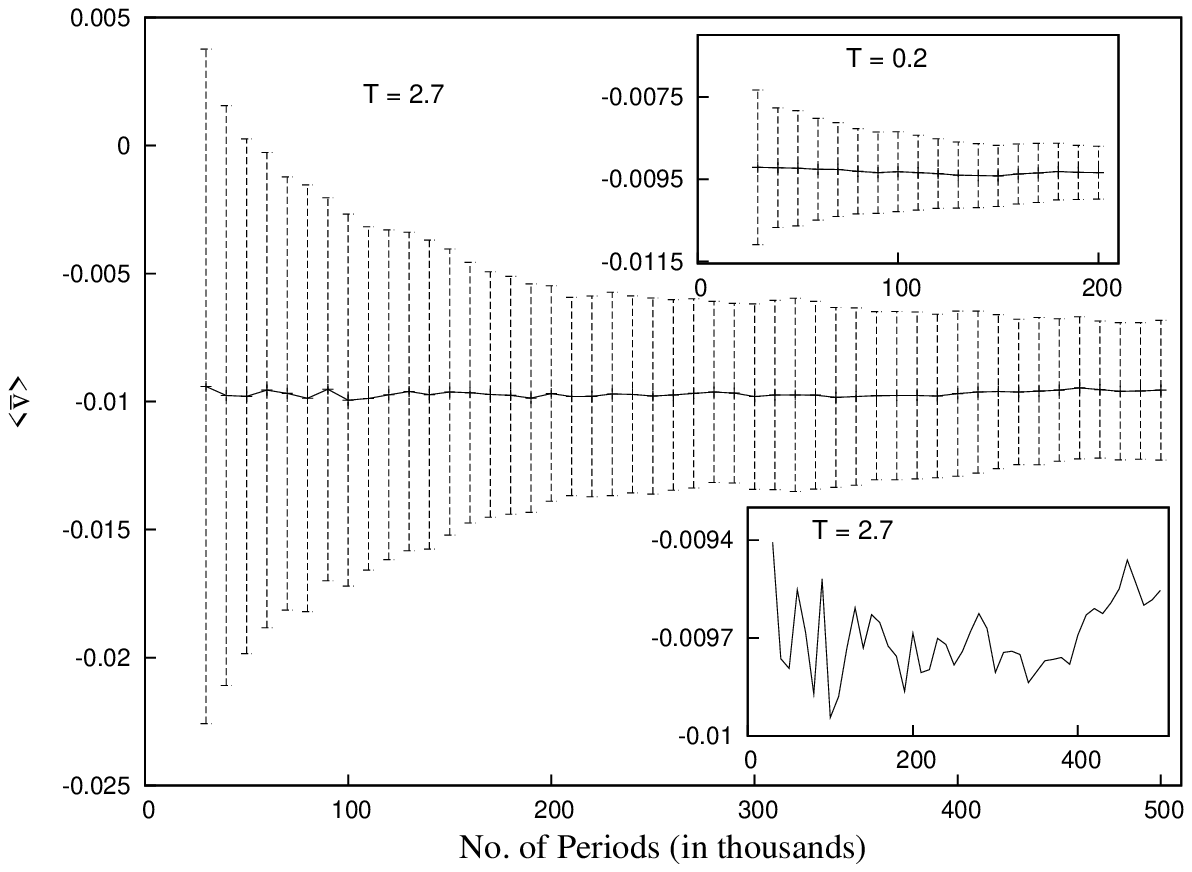}
\caption{Variation of ${<\overline{v}>}$ as a function of the number of 
periods of the forcing used to calculate ${\overline{v}}$'s. The data for 
${<\overline{v}>}$ had been recorded for every 10000 cycles and the 
corresponding error bars are calculated. Here, $\tau$ = 7.7, 
$\theta$ = $0.5\pi$, $\gamma_0 = 0.07$, $\lambda = 0.9$, $T = 2.7$ and
$F_0 = 0.2$. The inset in the bottom-right shows the variation of
${<\overline{v}>}$ without error bars. The inset in the top-right corner is 
same as the main graph but for $T$=0.2. (Note: The labels on the axes of the 
insets are same as those of the main graph).}
\end{figure}

\begin{figure}[htp]
\centering
\includegraphics[width=14cm,height=8cm]{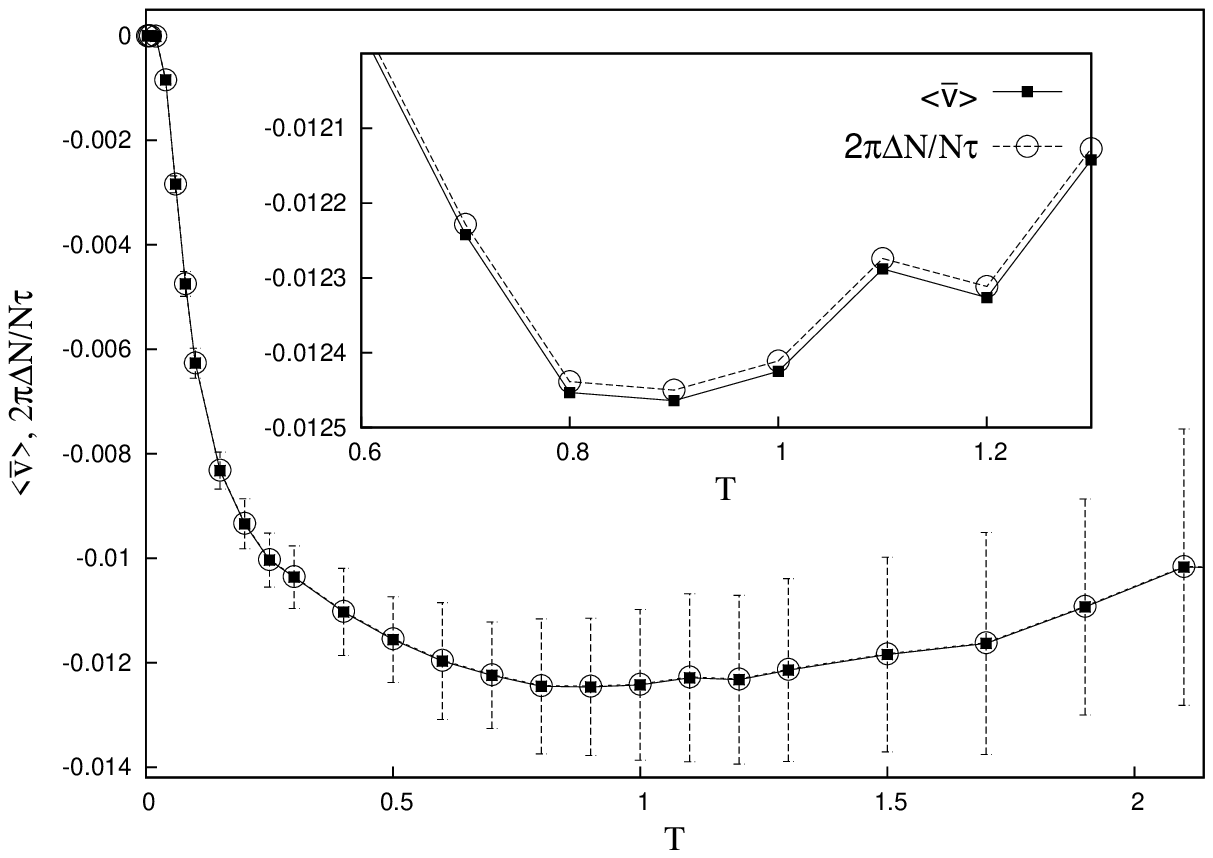}
\vspace*{-8mm}
\caption{Variation of ${<\overline{v}>}$ as a function of T is shown (dark 
squares with errorbars) for $\tau$ = 7.7, $\theta$ = $0.5\pi$, 
$\gamma_0 = 0.07$, $\lambda = 0.9$, and $F_0 = 0.2$. In the graph are also 
included the plot (open circles) of $\frac{2\pi\Delta N}{N\tau}$, where 
$\Delta N$ is the difference between the number of left and right interwell 
hoppings, and $N$ is the corresponding total number of periods of $F(t)$ 
considered. The inset shows the magnified image of the main plot (without the 
errorbars) for a limited range of T values showing that both the quantities 
${<\overline{v}>}$ and $\frac{2\pi\Delta N}{N\tau}$ are very close but not 
exactly equal.}
\end{figure}

At large temperatures the effect of frictional inhomogeneity gets manifest. 
However, one needs to calculate the mean values carefully. The frictional 
inhomogeneity provides only a weak (left-right) asymmetry. As a consequence, 
the ratchet current $<\overline{v}>$ obtained is not large and the standard 
deviations, $\Delta v$ (errorbars), are not small, though smaller than the 
mean $<\overline{v}>$. This is especially true when the temperatures are 
large. Therefore, whereas for smaller temperatures $T<0.5$ one gets sensible 
$<\overline{v}>$, values, say $\frac{\Delta v}{<\overline{v}>} < 0.2$, when 
averaged over 200000 periods of the drive, it takes 500000 periods at 
temperatures $T>1.0$ to obtain similar sensible $<\overline{v}>$ values. 
Figs. 2 and 3 illustrate the above statement succinctly.

In Fig.3 is also plotted $\frac{2\pi\Delta N}{N\tau}$ as a function of 
temperature. Here $\Delta N$ is the difference between the number of 
transitions to the left and right direction potential wells and $N$ is the
total number of periods of $F(t)$ used to obtain $\Delta N$. The figure shows
that $<\overline{v}>$ and $\frac{2\pi\Delta N}{N\tau}$ almost coincide. Note 
that the possibility of the particle making transition to a potential well 
farther than the adjacent well of the starting well before getting trapped is 
not ruled out, though such events are rare at lower temperatures but not so 
rare at higher temperatures. (These events are actually observed in the 
trajectory plots.) This naturally brings in the question of distribution of
jump lengths (in terms of number of wells) on either directions and hence the
dependence of net displacements (and ratchet currents) on the distributions.
However, from the matching of $<\overline{v}>$ and $\frac{2\pi\Delta N}{N\tau}$
in Fig.3 shows that the disribution is sharply peaked at the jump length of
$2\pi$, or one well-length, on either directions; distributions are almost 
identical on both the directions. Of course, a finite small tail of the 
distributions corresponding to larger lengths, at least at higher temperatures,
is always there.

Of course, clearly, the quantities of our interest $<\overline{W}>$ and 
$<\overline{v}>$ depend on the parameters $F_0$ and $\omega=\frac{2\pi}{\tau}$ 
of the external field $F(t)$. In what follows we shall keep the amplitude 
$F_0$ fixed and equal to 0.2 throughout. The chosen value of $F_0$ is not only 
because it corresponds to a subthreshold forcing but it is in a range where 
one is expected to get desired sensible results\cite{Wanda}.

\subsection{Dynamical states of trajectories at low temperatures}

\begin{figure}[htp]
\centering
\includegraphics[width=14cm,height=7cm]{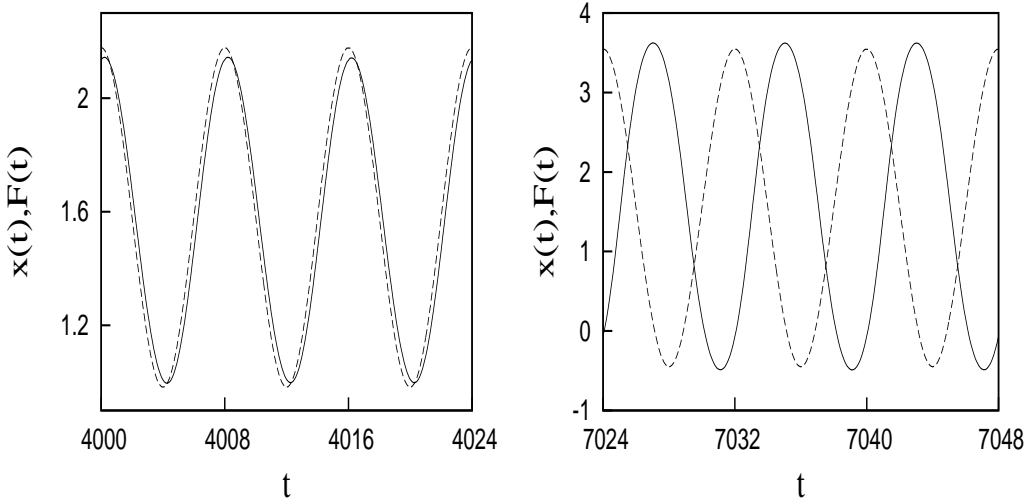}
\vspace*{-8mm}
\caption{The figure shows the two dynamical states of trajectories: SA state 
(left), with x(0) = 0.40$\pi$ and LA state (right), with x(0) = -0.35$\pi$, 
occuring when driven by a cosinusoid of same amplitude $F_0$ = 0.2 at 
$T = 0.000001$, with $\lambda$ = 0.9, $\gamma_0$ = 0.08, $\theta$ = 0.50$\pi$ 
and $\tau$ = 8.0. For both the figures, the dashed lines represent $F(t)$ 
multiplied by 2.99 centred at 1.58 and by 10 centred at 1.55, respectively, 
for easy comparison.}
\end{figure}

Before we begin presenting the numerical results at finite temperatures, we 
examine the trajectories at a temperature close to zero ($T=0.000001$) as a 
response to an external field $F(t)$. The nonuniformity of the friction 
coefficient $\gamma(x)=\gamma_0(1-\lambda\sin(x+\theta))$ is measured by the
value of $\lambda$ and its phase shift with respect to $V(x)=-\sin(x)$
by $\theta$. Fig.4 shows typical trajectories in the two states of large
amplitude (LA) and small amplitude (SA) depending on the initial conditions
chosen but for the same $F(t)$. These trajectories are periodic as $F(t)$ and 
are characterized by amplitude $x_0$ and phase lag (with respect to $F(t)$) 
$\phi$. Since, the dynamics being stochastic, the amplitude and phase lag are 
not the same in every period of $F(T)$, only their average values make sense.
We measure them in either of the two ways: (i) we calculate the mean values
of the quantities over several periods from the trajectories, or (ii) 
calculate from the mean hysteresis loops, Fig.1, considering this 
approximately to be an ellipse. However, the second method becomes impractical 
when the hysteresis loop becomes non-elliptic and in that case the first
method is the only option.

\begin{figure}[htp]
\centering
\includegraphics[width=15cm,height=17cm]{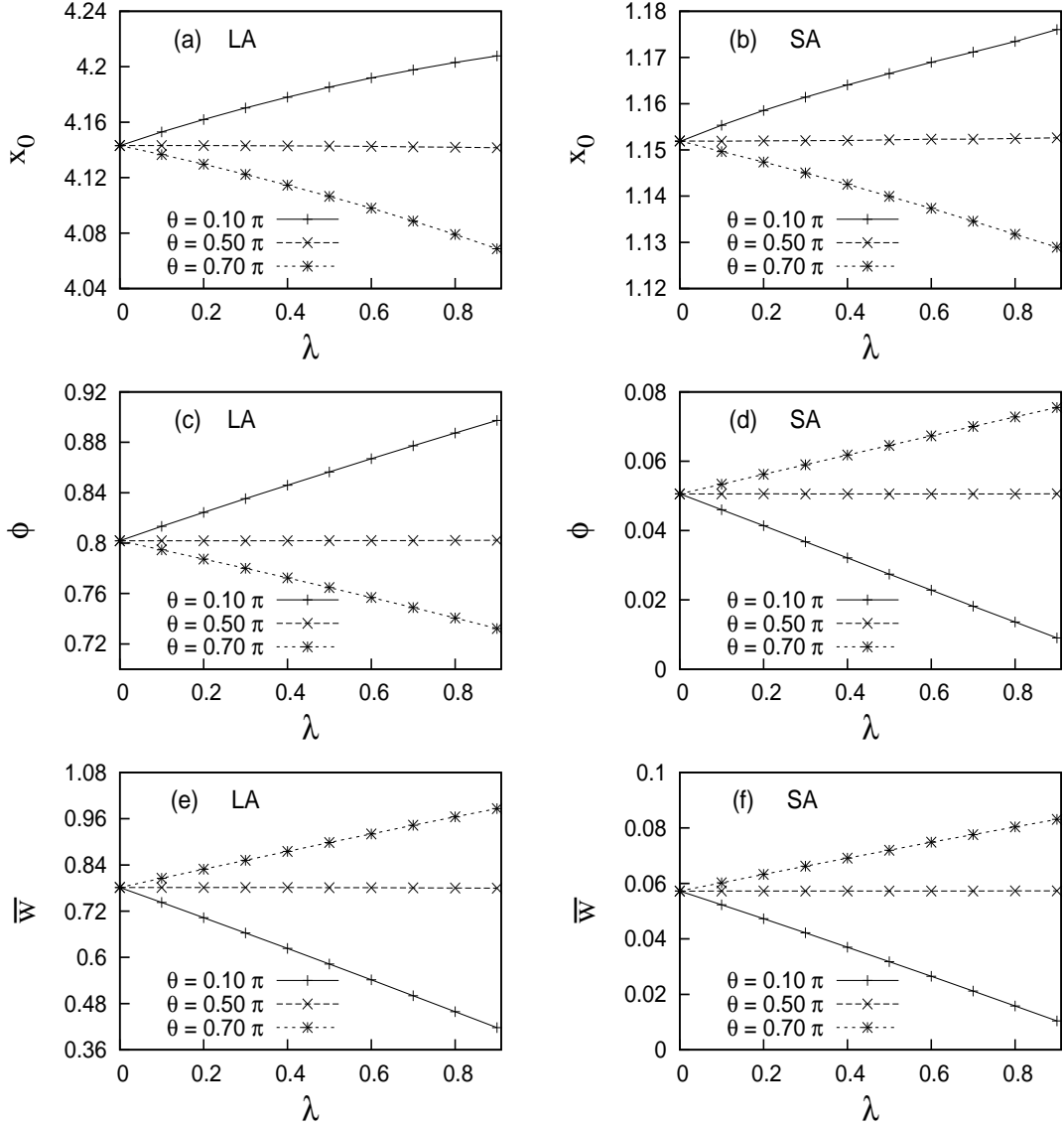}
\vspace*{-20mm}
\caption{The variations of $x_0$, $\phi$ and $\overline{W}$ as a function of 
$\lambda$ for both LA and SA states with $F_0$ = 0.2, $\tau$ = 8.0, $\gamma_0$ 
= 0.07 at $T$ = 0.000001 and various values of $\theta$ as indicated.}
\end{figure}

In Figs.5, the variation of $x_0$ is shown as a function of $\lambda$ for 
various values of $\theta$ in the two states: LA state (a) and SA state (b). 
The percentages of variations are calculated with reference to values at 
$\lambda=0$. In Figs. 5(c) and 5(d), the variations of $\phi$ are shown. It is 
to be noted, from the figures, that for $\theta=\frac{\pi}{2}$ both $x_0$ and 
$\phi$ almost remain constant (variation less than 0.03\%) over the entire 
range of $0<\lambda<1$. For $\theta<\frac{\pi}{2}$, $x_0$ increases as a 
function of $\lambda$ whereas for $\theta>\frac{\pi}{2}$, $x_0$ decreases with 
increase of $\lambda$ for both LA and SA states. However, these variations are 
very small, less than 2\% at $\lambda=.9$. In the SA state the values of 
$\phi$ themselves are very small and hence their variations, though very 
small, appear to be large in percentage. In this state, for 
$\theta<\frac{\pi}{2}$, $\phi$  decreases and for $\theta>\frac{\pi}{2}$, it 
increases as a function of $\lambda$. The LA state has large values of $\phi$, 
for example, $\phi\sim 0.8\pi$ at $\theta=\frac{\pi}{2}$. In this LA state, as 
$\lambda$ is increased from a small value ($\lambda\approx 0$) to a large 
value, $\phi$ increases for $\theta<\frac{\pi}{2}$ whereas it decreases for 
$\theta>\frac{\pi}{2}$. These variations are comparatively large ($\leq 10$\%).
Since the variations of $x_0$ are small the variations of the hysteresis 
($\overline{x(F)}$) loop area $\overline{A}$ (or the input energy 
$\overline{W}$) in the two states are determined essentially by the variations 
of $\phi$. Note that $\phi=\frac{\pi}{2}$ should have the largest 
$\overline{W}$ for a given $x_0$. These varations of $\overline{W}$ are shown 
in Figs. 5(e) and 5(f).

\begin{figure}[htp]
\centering
\includegraphics[width=15cm,height=15cm]{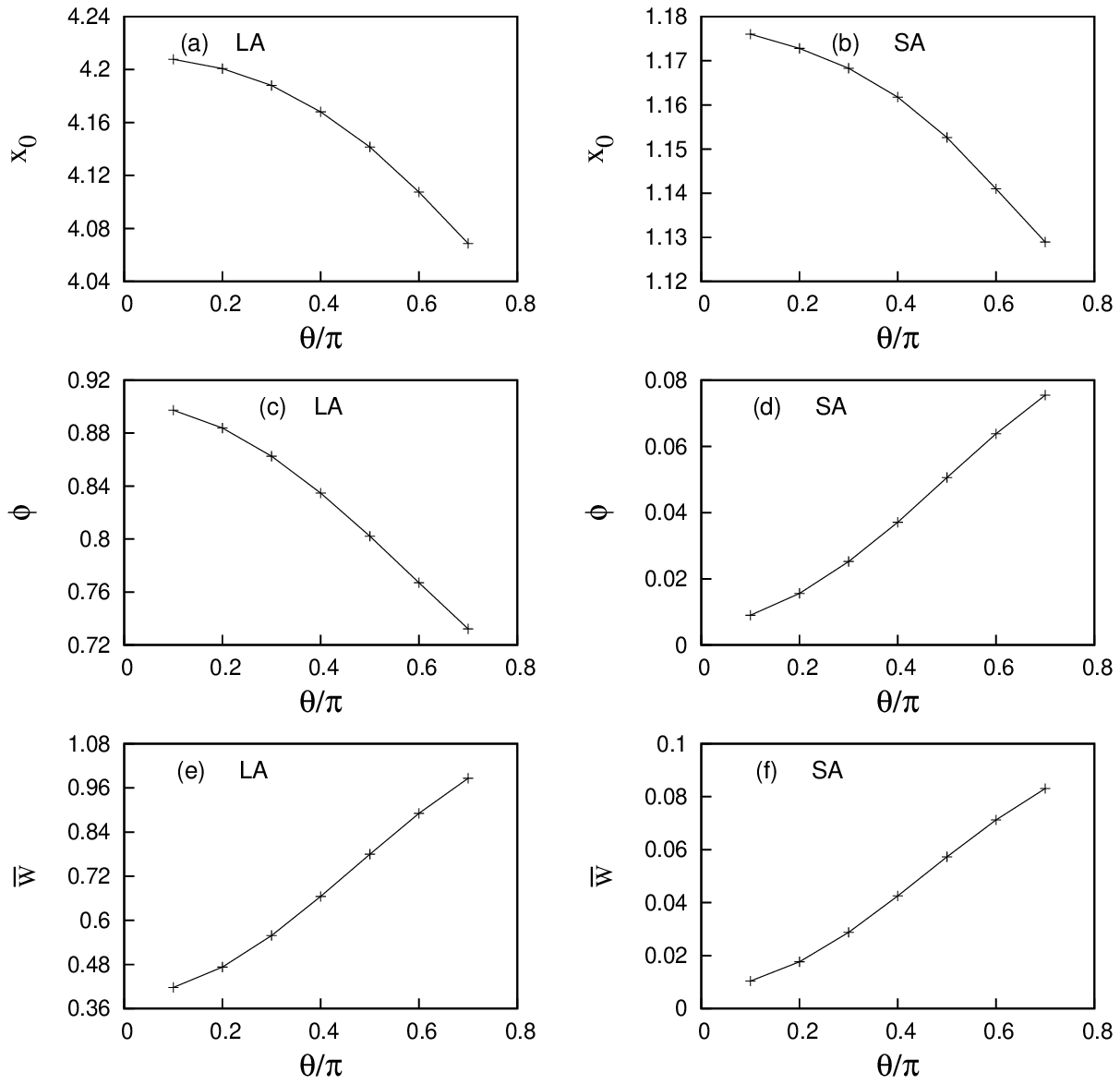}
\vspace*{-20mm}
\caption{The variations of $x_0$, $\phi$ and $\overline{W}$ as a function of 
$\theta$ for both LA and SA states with $F_0$ = 0.2, $\tau$ = 8.0, $\gamma_0$ 
= 0.07 at T = 0.000001, keeping $\lambda$ = 0.9 fixed.}
\end{figure}
 
In what follows we shall keep the value of $\lambda=0.9$ fixed. For 
$\lambda=0.9$ the results of Figs. 5 are summarized in Figs. 6(a-f):  
$x_0(\theta)$, $\phi(\theta)$, and $\overline{W}(\theta)$ for both LA and SA
states. Again, the variation of $\overline{W}(\theta)$ is determined by the 
variation of $\phi(\theta)$; for the LA state $\phi$ decreases towards 
$\frac{\pi}{2}$ (and hence $\overline{W}$ increases), whereas for the SA state
$\phi$ increases as $\theta$ is increased 
(again thereby increasing $\overline{W}$). Note that, since the particle moves 
in a periodic potential with a potential barrier of magnitude 2 whereas the 
drive amplitude is only 0.2 (and hence maximum reduction in potential barrier 
= $0.4\pi$) the particle will remain confined to the same well of the potential 
at low temperatures. Only when the temperature is increased that the particle 
will have any chance of overcoming the potential barrier. Therefore, the mean 
net velocity $<\overline{v}>$ is zero at low temperatures even though 
$<\overline{W}>$ is nonzero at all temperatures.

\begin{figure}[htp]
\centering
\includegraphics[width=15cm,height=10cm]{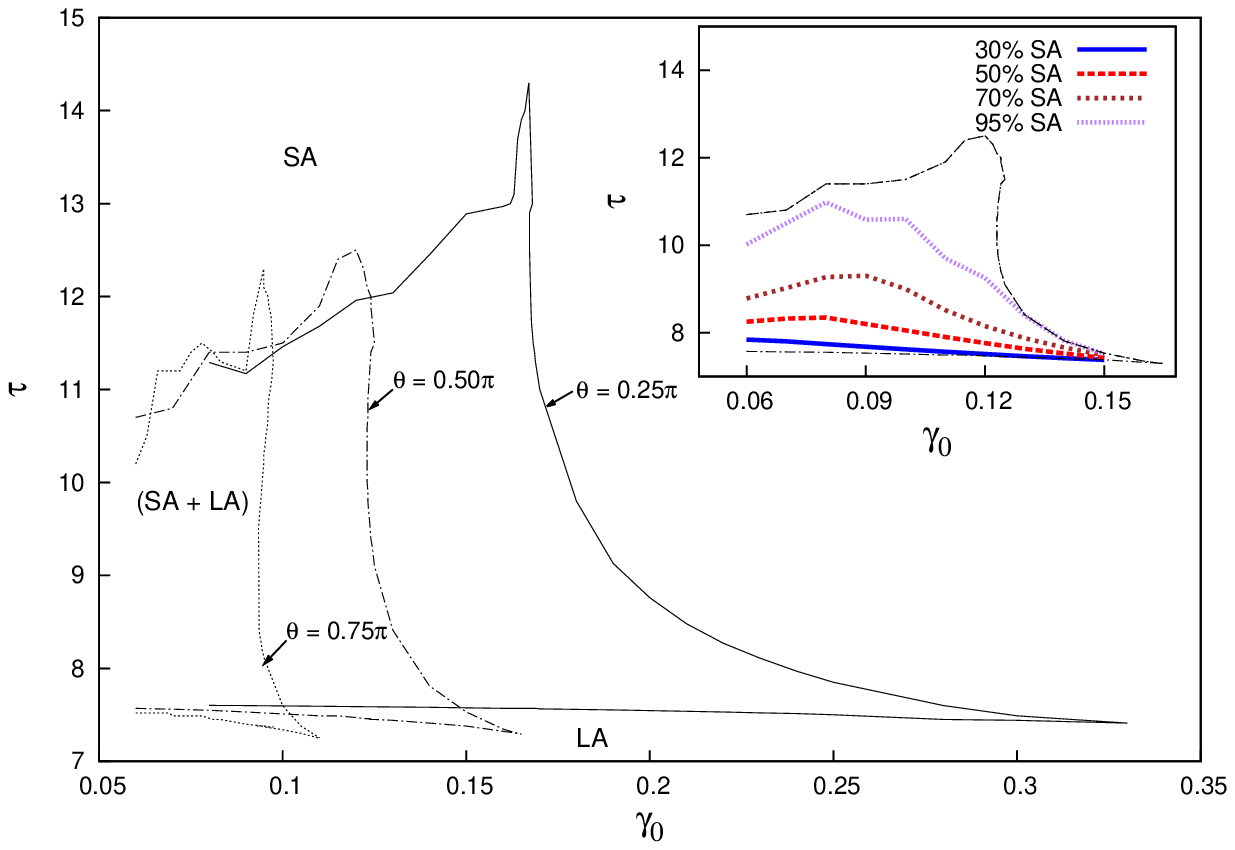}
\vspace*{-5mm}
\caption{The figure shows the boundaries of regions of coexistence of SA and LA
states for $\theta=0.25\pi,~0.5\pi$ and $0.75\pi$ at $T=0.000001$. Observe 
that $\theta$ = 0.25$\pi$ provides a larger maximum limiting value of 
$\gamma_0$ in the ($\tau$,$\gamma_0$) space for the coexistence of the 
dynamical states. Clearly, with $\theta$ = 0.75$\pi$, the limiting value of
$\gamma_0$ has reduced significantly. In the inset are shown the loci of 
various percentages of populations of the SA state for $\theta$ = 0.5$\pi$ at 
T= 0.000001.}
\end{figure}

Figure 7 shows the coexistence boundary of the two dynamical states in the
$(\gamma_0,\tau)$ space for $\theta=0.25\pi,~0.5\pi$, and $0.75\pi$. In the 
inset of the figure are shown the loci of various fractions of SA state 
within the coexistence boundary for $\theta=0.5\pi$. Similar loci for other
values of $\theta$ have also been calculated but are not shown here. However, 
the consequences of such fractions at any point in the $(\gamma_0,\tau)$ space
will be discussed later. Clearly, the upper limit of $\gamma_0$ for the 
coexistence region decreases as $\theta$ increases. This behaviour of $\theta$ 
dependence can be understood if the relationship between the potential $V(x)$ 
and $\gamma(x)$ is examined. It will be easier to visualize if a rectangular 
profile of $\gamma(x)$ is considered, for illustration, instead of a 
sinusoidal profile. This is justified because it has been found that 
rectangular and sinusoidal profile of $\gamma(x)$ yield essentially similar 
transport characteristics.

\begin{figure}[htp]
\centering
\includegraphics[width=15cm,height=7cm]{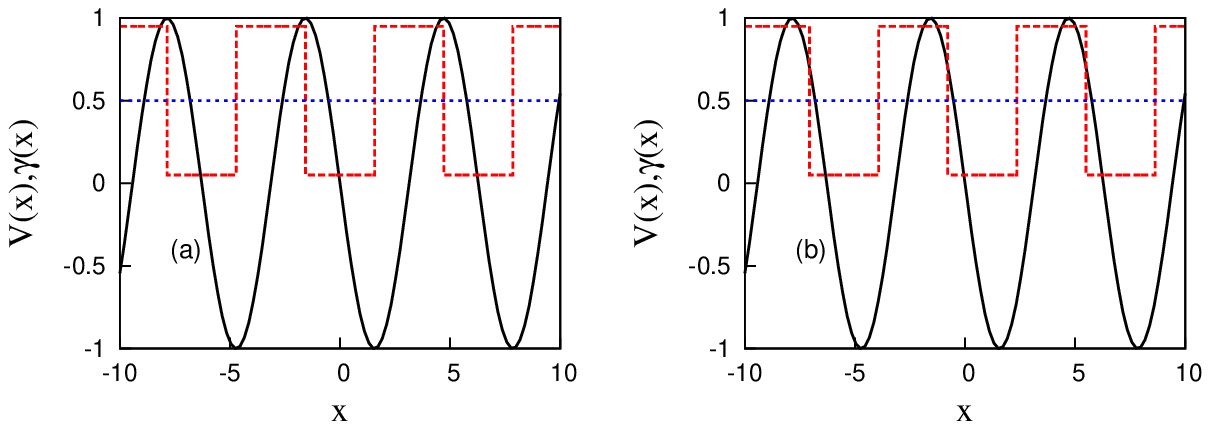}
\vspace*{-5mm}
\caption{The above two figures compare the the potential V(x) (black) and the
(square-wave type) friction coefficient $\gamma$(x) (red). The left figure is 
for $\theta$ = 0.50$\pi$ and the right figure is for $\theta$ = 0.25$\pi$. The 
blue line in both the figures is the mean value of $\gamma$(x), here, taken as 
$\gamma_0$ = 0.5 for better comparison.}
\end{figure}

In Fig.8, as an illustration, the potential $V(x)$ and rectangular $\gamma(x)$ 
are plotted together 
for the phase difference $\theta=0.5\pi$ (a) and $\theta=0.25\pi$ (b). Since 
$V(x)$ is sinusoidal the space between two consecutive minima forms a period. 
From Fig.8(a) one can see that $\gamma(x)$ is large over the entire half period 
to  the left of $V(x)$ peak whereas $\gamma(x)$ is small over the rest half 
period, to the right, of $V(x)$ peak. Therefore, a particle will experience a 
moderate effective damping force, of magnitude, say $F_{eff}$, measured 
roughly by the averaged product $\gamma(x)v(x(t))$ over a period. On the other 
hand, from Fig.8(b), for $\theta=0.25\pi$, one can see that at the bottom of 
the potential, where the particle velocity is expected to be large, the 
friction coefficient $\gamma$ is small and at the peak of the potential, where 
the velocity should be small the friction is large. Thus, one can conclude that 
the effective damping force experienced by the particle in this case will be 
smaller than $F_{eff}$ of the former case and hence the particle will be more 
mobile in this case of smaller $\theta$. From a similar consideration one can 
see that the effective damping force experienced by the particle in case of, 
say, $\theta=0.75\pi$ will be larger than $F_{eff}$. In other words, the 
average mobility of the particle decreases as a function of $\theta$ for the
same forcing and the potential function. Though difficult to define, one can 
think of a uniform effective friction coefficient $\overline{\gamma_0}$ for 
each $\theta$ (for nonuniform $\gamma(x)$) for a given potential function and
external field, as in case of ref.\cite{Saikia}. Thus, one can argue that the 
average effective friction coefficient, $\overline{\gamma_0}$ increases 
monotonically as $\theta$ is changed from 0 to 
$\pi$ in $\gamma(x)=\gamma_0(1-\lambda\sin(x+\theta))$. This explains why the 
($\gamma_0,\tau$) region of coexistence of the two dynamical states in Fig.7 
shrinks as $\theta$ increases in order to have the same effective $\gamma$ at 
the upper coexistence limit.

\subsection{Ratchet current and input energy at finite temperatures}

\begin{figure}[htp]
\centering
\includegraphics[width=14cm,height=10cm]{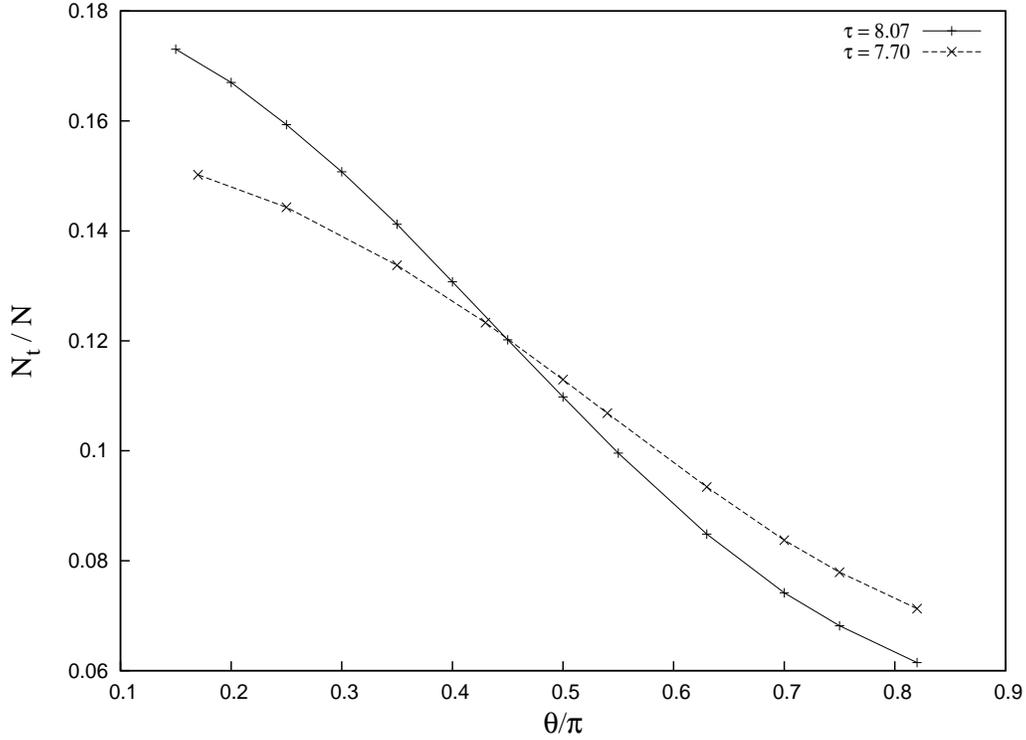}
\vspace*{-5mm}
\caption{The figure shows the decrease of the fraction of the total number of 
interwell hoppings ($\frac{N_t}{N}$) as a function of $\theta$, for two 
$\tau$ values, $\tau$ = 8.07 (upper plot) and $\tau$ = 7.70 (bottom plot) in 
200000 periods of $F(t)$. Here, $T = 0.2$, $F_0$ = 0.2, $\gamma_0$ = 0.07, and 
$\lambda$ = 0.9.}
\end{figure}

From the above analysis one can conclude that for a given pair of 
($\gamma_0,\tau$) values the effective friction coefficient 
$\overline{\gamma_0}$ increases with $\theta$ and hence, as can be seen in 
Fig.9, the total number ($N_t$) of interwell transitions at a given 
temperature $T$ decreases monotonically. This, however, does in no way imply 
that the ratchet current decreases monotonically with increasing $\theta$. 

\begin{figure}[htp]
\centering
\includegraphics[width=14cm,height=10cm]{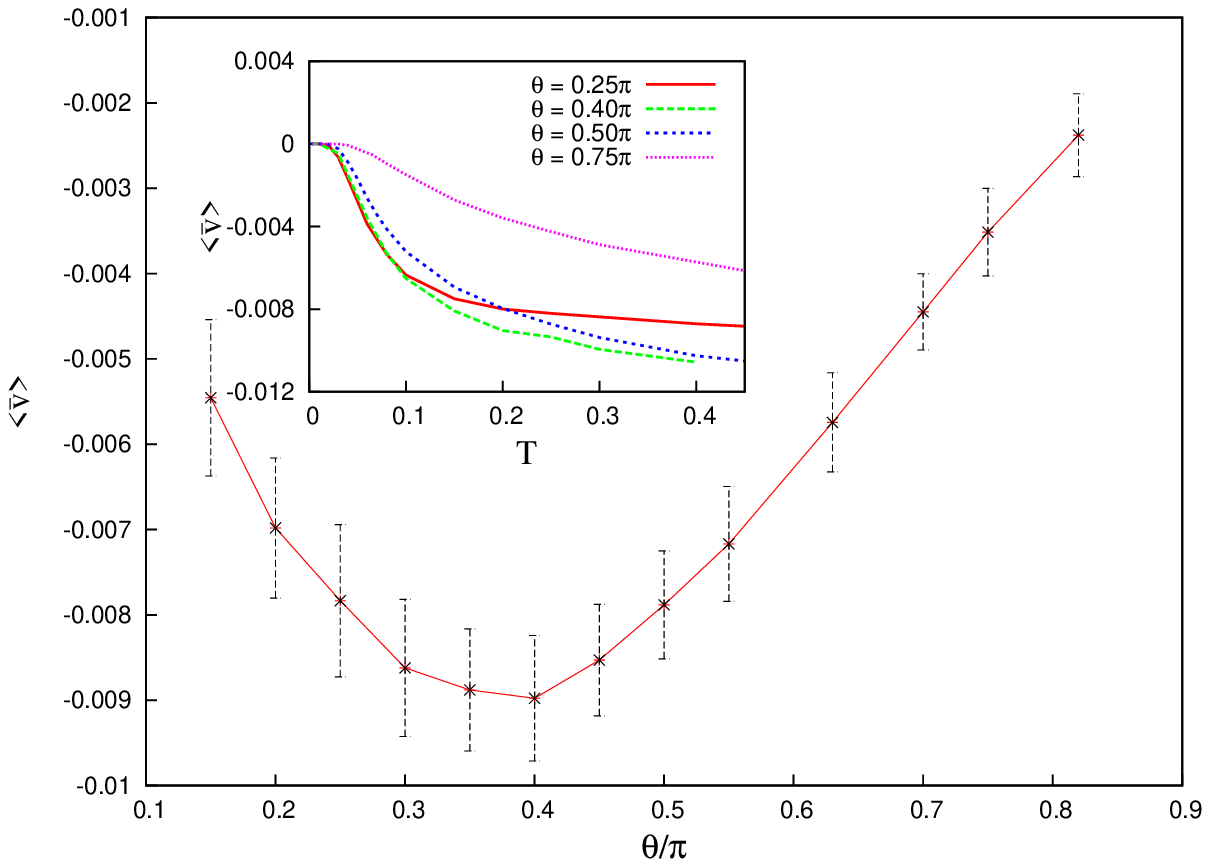}
\vspace*{-5mm}
\caption{Variation of ${<\overline{v}>}$ with $\theta$ at T = 0.2 for $\tau$ = 
8.07, $\lambda$ = 0.9 and $F_0$ = 0.2. The inset shows the variation of 
${<\overline{v}>}$ with T for $\tau$ = 8.07 but for different $\theta$ values 
as indicated in the plot keeping other parameters same.}
\end{figure}

As can be seen, in Fig.10, the ratchet current, which is also a measure of 
difference between the interwell transitions in the right and left directions, 
peaks around $\theta=0.5\pi$. Since $\theta=0.5\pi$ provides the largest 
frictional asymmetry one would ideally think of the ratchet current to be 
maximum at $\theta=0.5\pi$. However, at lower temperatures the ratchet current 
actually peaks at $\theta<0.5\pi$, Fig.10. This is because lower $\theta$ 
corresponds to lower effective $\overline{\gamma_0}$ and hence being more 
mobile interwell transitions begin at lower temperatures than for larger 
$\theta$. Thus ratchet current grows faster, beginning at lower temperatures, 
for smaller $\theta$ and hence dominate before the frictional asymmetry 
overwhelms it at larger temperatures where one can find ratchet current to be 
the maximum for $\theta=0.5\pi$, inset of Fig.10.

\begin{figure}[htp]
\centering
\includegraphics[width=15cm,height=10cm]{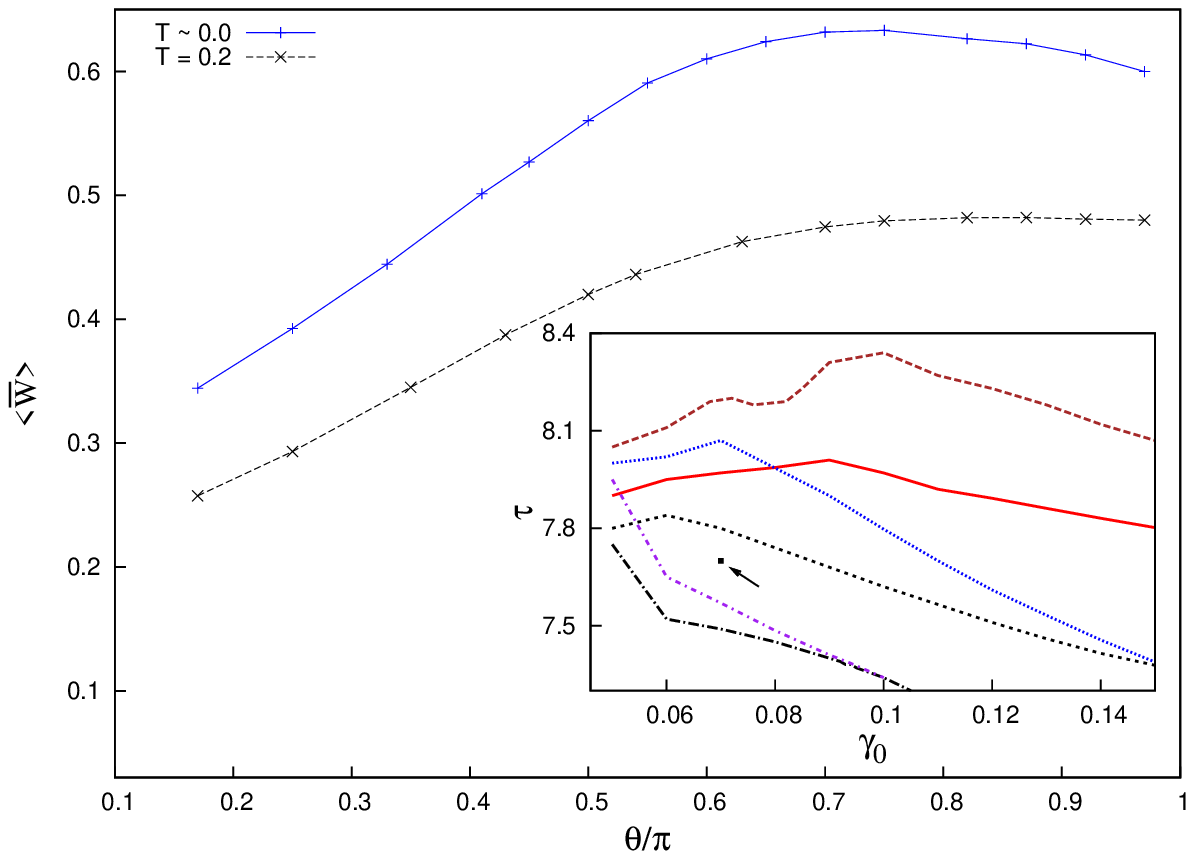}
\vspace*{-5mm}
\caption{The variation of $<\overline{W}>$ as a function of $\theta$ is shown 
for $\tau$ = 7.70, $F_0$ = 0.2, $\lambda$ = 0.9, $\gamma_0$ = 0.07 at 
$T=0.000001$ (blue) and $T = 0.2$ (black). The inset shows the loci of 
constant percentages of population of the SA state in the ($\tau$-$\gamma$) 
space at $T=0.000001 $ for various $\theta$ values: 30\% population of SA 
states when $\theta$ = 0.25$\pi$, $\theta$ = 0.50$\pi$ and $\theta$ = 0.75$\pi$
are, respectively, depicted by red, black dotted and black dash-dotted lines, 
whereas for 40\% population of SA states are correspondingly shown in brown, 
blue and purple. The point ($\gamma_0$=0.07,$\tau$=7.7) is marked in the 
inset.}
\end{figure}

In Fig.11, the input energy $<\overline{W}>$ is plotted as a function of
$\theta$ for two temperatures, (i) $T=0.000001$, based on the weightage 
average of results for LA and SA states (Figs. 6(e,f)), and (ii) $T=0.2$, for 
$\gamma_0=.07$ and $\tau=7.7$. At low temperatures $<\overline{W}>(\theta)$ 
strongly depends on the choice of the point $(\gamma_0,\tau)$ and the variation
can be explained with the help of the coexistence diagrams, Fig.7 and the loci
of fractions of the SA states. In the inset of Fig. 11 are drawn some 
representative loci of the fractions of the SA state for 
$\theta=0.25\pi,~0.5\pi$, and $0.75\pi$ together around the point (0.07,7.7) 
in the $(\gamma_0,\tau)$ space. One can find that the fractions of the
SA states for $\theta=0.75\pi$ is 38.5\%, for $\theta=0.5\pi$
it is 25\% and for $\theta=0.25\pi$ it is 17.5\%. In other words, the 
fractions of the LA states are respectively, 61.5\%, 75\%, and 82.5\% for
$\theta=0.75\pi,~0.5\pi$, and $0.25\pi$. Since $\overline{W}$ for LA states are 
much larger compared to $\overline{W}$ for SA states the weightage average
$<\overline{W}>$ decreases with $\theta$ because the fraction of LA states 
decreases with $\theta$. However, the $\overline{W}$ for LA and SA states are 
monotonically increasing as a function of $\theta$, Fig.6, implying thereby 
that $<\overline{W}>$ increases with $\theta$ if the fractions are kept fixed. 
These two opposing trends of variations of $<\overline{W}>$ as a function of 
$\theta$ leads to the peaking behaviour of $<\overline{W}>$ as a function of 
$\theta$ at low temperatures. However, as the temperature is increased the 
transitions between the LA and SA states allow SA states to appear where 
previously only LA states existed and vice versa. This makes the coexistence 
boundaries of Fig.7 blurred and nonexistent and hence $<\overline{W}>(\theta)$ 
gets smoothened. Fig.11(b) shows that, at $T=0.2$, $<\overline{W}>(\theta)$ is 
monotonic which is to be compared with $<\overline{v}>$, shown in Fig.10, 
showing a peak close to $\theta=\frac{\pi}{2}$ irrespective of the choice of 
$(\gamma_0,\tau)$ point. Thus, there is no correlation between energy 
absorption and the ratchet current, at least as far as variation with respect 
to $\theta$ is concerned.

\begin{figure}[htp]
\centering
\includegraphics[width=14cm,height=8cm]{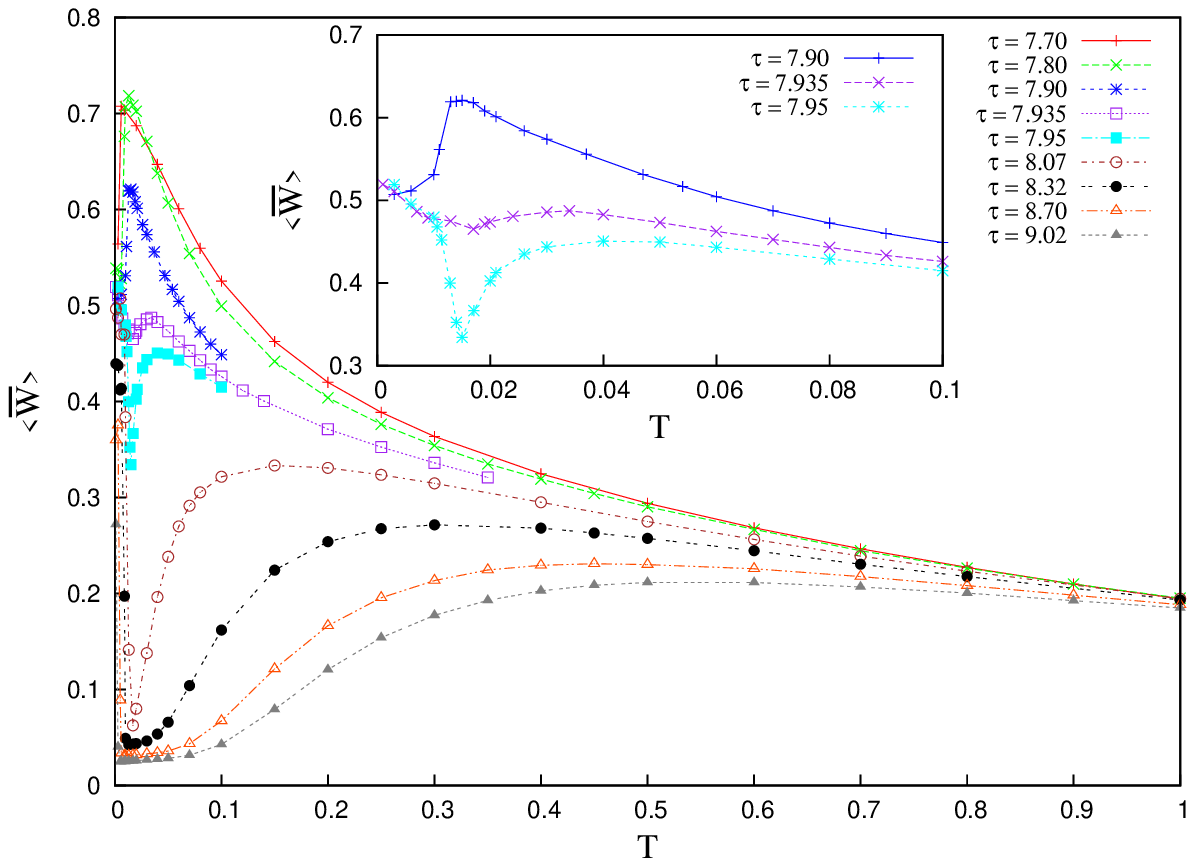}
\vspace*{-4mm}
\caption{Variation of $<\overline{W}>$ as a function of $T$ for various $\tau$ 
values as indicated in the top-right corner of the figure with $F_0$ = 0.2, 
$\gamma_0$ = 0.07, $\lambda$ = 0.9, $\theta$ = 0.50$\pi$. The inset displays 
the two differing nature of input energy variations at low temperatures with 
$\tau\approx$ 7.935 as the boundary.}
\end{figure}

\begin{figure}[htp]
\centering
\includegraphics[width=14cm,height=8cm]{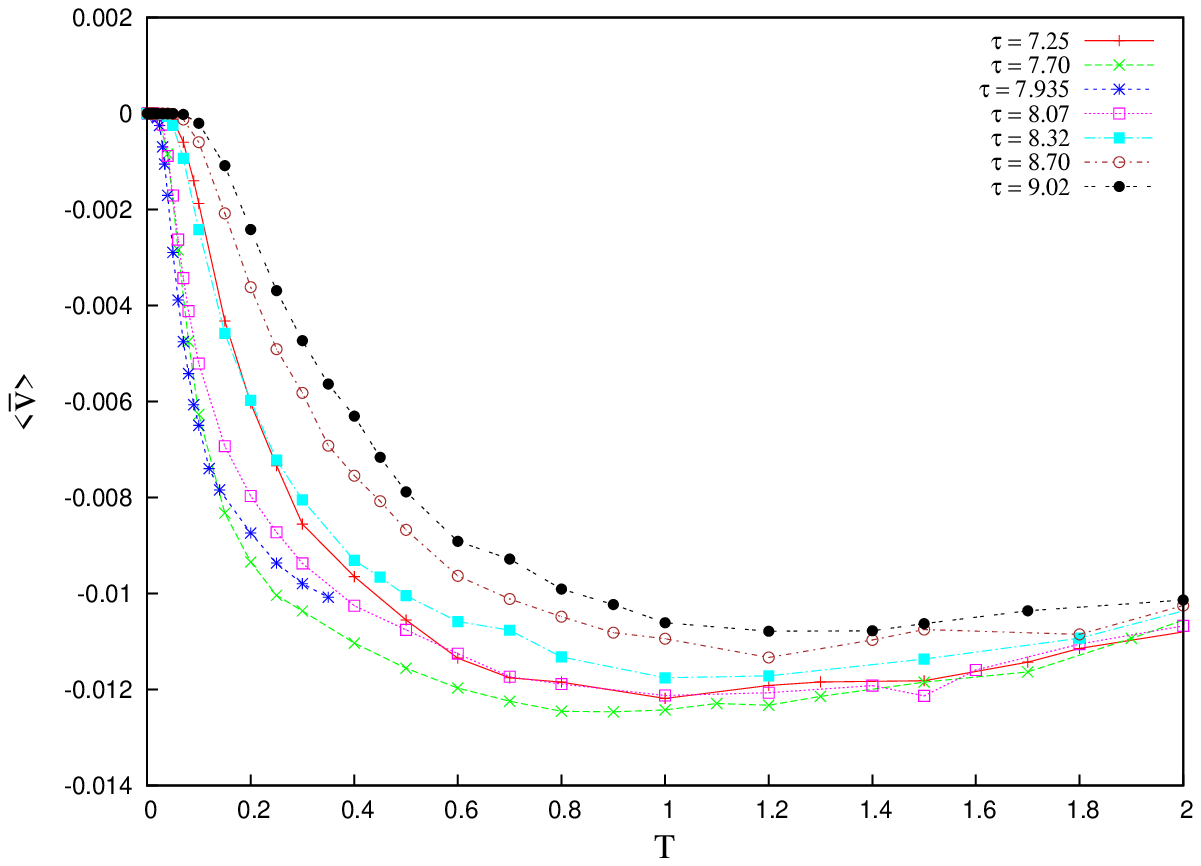}
\vspace*{-4mm}
\caption{Variation of $<\overline{v}>$ as a function of $T$ for various $\tau$ 
values as indicated in the top-right corner of the figure with $F_0$ = 0.2, 
$\gamma_0$ = 0.07, $\lambda$ = 0.9, $\theta$ = 0.50$\pi$.}
\end{figure}

Figure 7 also provides a guideline to choose the region of $(\gamma_0,\tau)$ 
space in order to obtain stochastic resonance for a given $\theta$. Invariably, 
stochastic resonance occurs if the region of $(\gamma_0,\tau)$ space is chosen 
such that both the dynamical states coexist, at low temperatures (for example,
$T=0.000001$), with their fractions not too far away from 0.5. In Fig.12, the 
mean energy absorbed $<\overline{W}>$ is plotted against temperature for 
various values of $\tau$ with $\gamma_0=0.07$. In Fig.13, the ratchet current 
$<\overline{v}>$ is plotted for the same parameters. From these figures one 
can observe that stochastic resonance  and ratchet effect do occur 
simultaneously in the same parameter space. Note, however, that the 
temperatures at which $<\overline{W}>$ and $<\overline{v}>$ peak are widely
separated; stochastic resonance and peak of ratchet current do not occur at 
close by temperatures. Similar results have also been obtained earlier when a 
homogeneous system was driven by a biharmonic forcing\cite{Wanda1}. Figs. 12 
and 13, obtained for an inhomogeneous system driven by a sinusoidal field, are 
the main results of the present work and deserve further analysis.

\begin{figure}[htp]
\centering
\includegraphics[width=14cm,height=10cm]{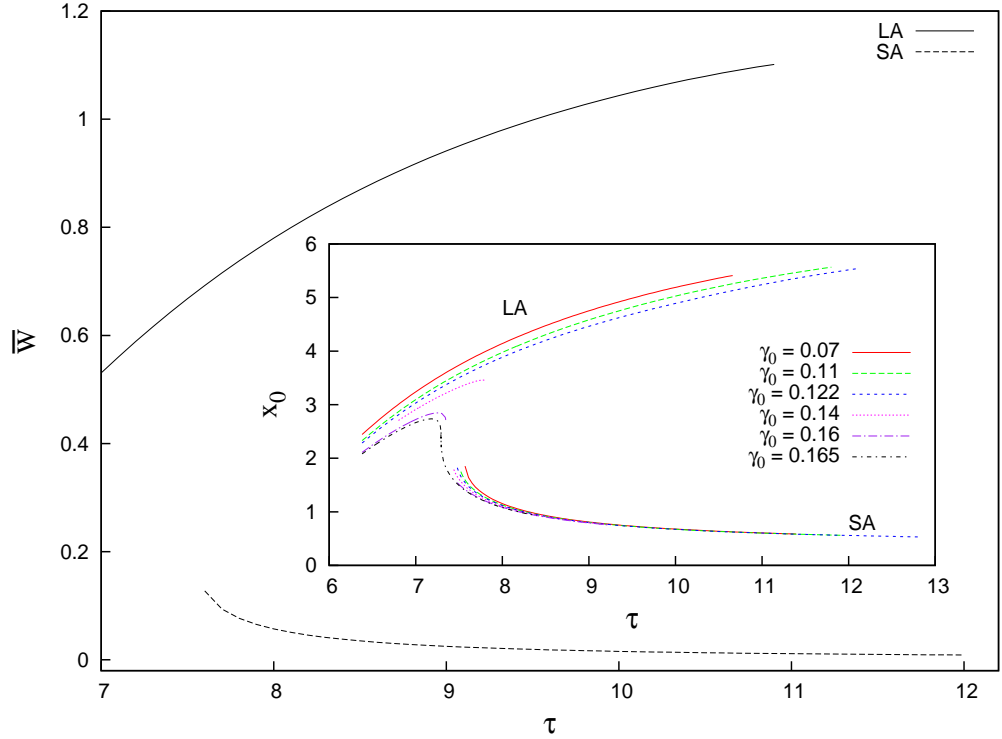}
\vspace*{-4mm}
\caption{Variation of $\overline{W}$ as a function of $\tau$ at $T=0.000001$ 
for the two dynamical (LA and SA) states as indicated, with $F_0$ = 0.2, 
$\gamma_0$ = 0.07, $\lambda$ = 0.9, $\theta$ = 0.50$\pi$. The inset shows the 
variation of amplitude $x_0$ as a function of $\tau$ for both LA and SA states
at $T=0.000001$, $\lambda$ = 0.9, $F_0$ = 0.2, $\theta$ = 0.5$\pi$ for 
different $\gamma_0$ values as indicated.}
\end{figure}

Figure 14 shows the variation of the input energy (or energy absorbed) per 
cycle, $\overline{W}$, of $F(t)$ as a function of the period $\tau$ of $F(t)$ 
in the LA and SA states at very low temperatures $T\approx 0$ and 
$\gamma_0=0.07$. Of course, the comparison makes sense only in the coexistence
region of ($\gamma_0,\tau$) space. The trend of variation of $\overline{W}$
qualitatively follows that of the amplitude $x_0$ of the trajectories, as
indicated by the curves in the inset of the figure. The trend remains the same 
as the temperature is increased by a small amount. Note that in the LA state 
the system absorbs more energy per cycle than when it is in the state SA. Of 
course, the mean input energy $<\overline{W}>$, as discussed earlier, is the 
weightage average of $\overline{W}$ in the two states. In view of these results
and based on the nature of variation of the curves in Fig. 12 at low 
temperatures, the set of curves in Fig.12 can be put in two qualitatively 
distinct groups: one roughly with $\tau\gtrsim 7.935$ and the other with 
$\tau\lesssim 7.935$, though the dividing line $\tau=7.935$ is not precise. The 
decrease of $<\overline{W}>$ due to thermal fluctuations as the temperature is 
raised from $T=0$ indicates transition from LA state to SA state. This shows 
that for the former group of curves with larger $\tau$ the dynamical (SA) state 
is more stable compared to the LA state. On the other hand, for the latter 
group of curves with lower $\tau$ the LA state is more stable compared to the 
SA state. This is consistent with the diagram shown in Fig.7. This is also 
understandable because larger is the frequency (smaller period) of drive 
larger should be the phase lag and hence the LA state (with larger phase lag) 
should be preferred and vice versa.

\section{Discussion and Conclusion}

We presented above the results of our calculations as a function of $T$ and
$\theta$ for various values of $\gamma_0$ and $\tau$. The results are 
consistent with each other and bear plausible explanations. However, some 
detailed results as a function of $\tau$ (or frequency, $\frac{2\pi}{\tau}$) 
need further scrutiny, which at present we lack clear explanation for. 

\begin{figure}[htp]
\centering
\includegraphics[width=14cm,height=20cm]{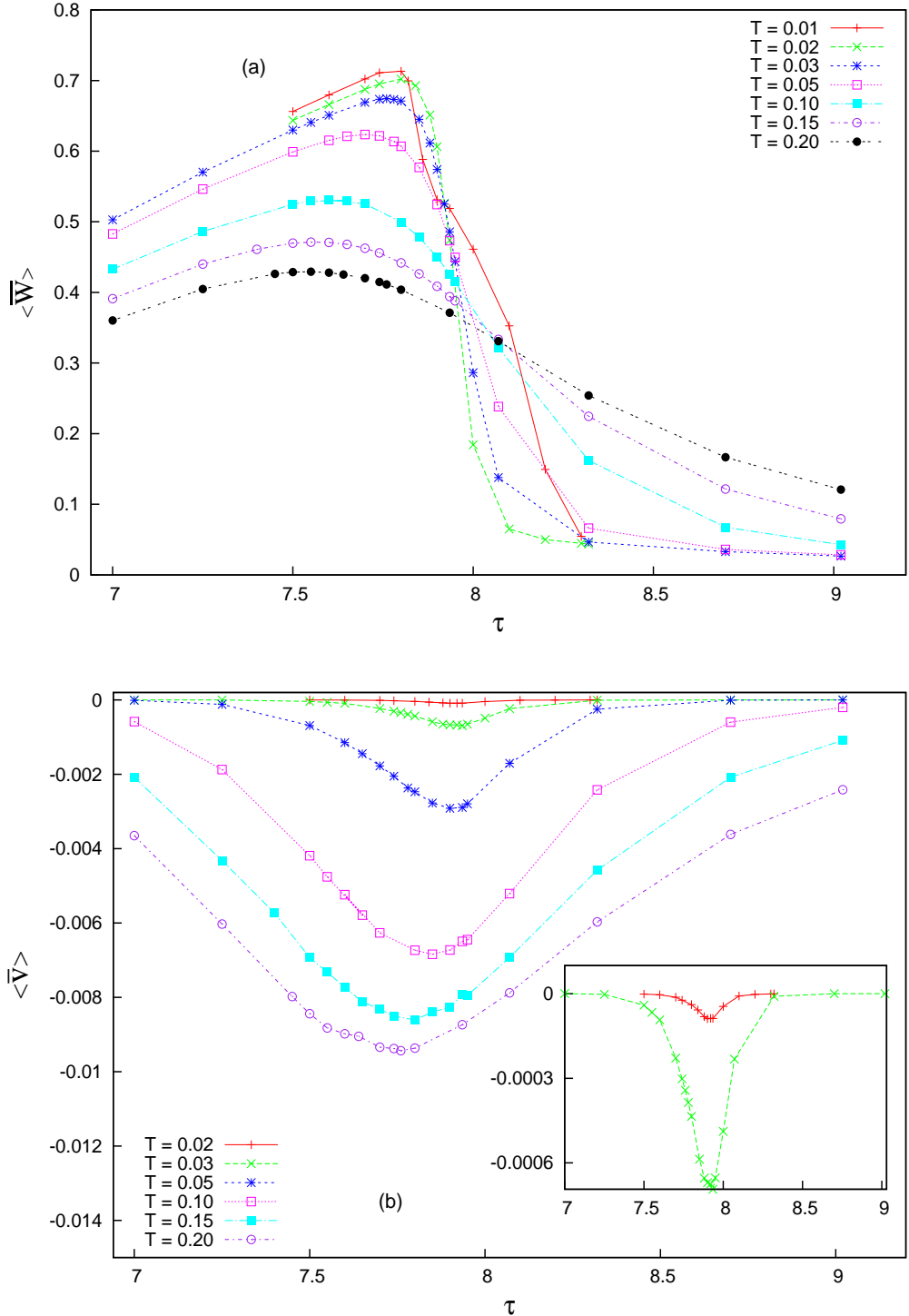}
\vspace*{-4mm}
\caption{Variation of $<\overline{W}>$ as a function of $\tau$ is shown in 
fig.15(a) and that of $<\overline{v}>$ in fig.15(b) for various small 
temperatures as indicated, with $F_0$ = 0.2, $\gamma_0$ = 0.07, $\lambda$ = 
0.9, $\theta$ = 0.50$\pi$. The inset in Fig.15(b), with the same axes label as 
the main plot, shows the two curves for lower temperatures $T=0.02$ and 
$T=0.03$ for clarity.}
\end{figure}

In Figs. 15 a and b are presented the variations of $<\overline{W}>$ and 
$<\overline{v}>$ as a function of $\tau$ for $\gamma_0=0.07$ and 
$\theta=0.5\pi$ at various temperatures. Particular attention may be given to 
the occurrence of maxima of $<\overline{W}>$ and $<\overline{v}>$ at low 
temperatures as summarized in Fig.16. 

\begin{figure}[htp]
\centering
\includegraphics[width=14cm,height=10cm]{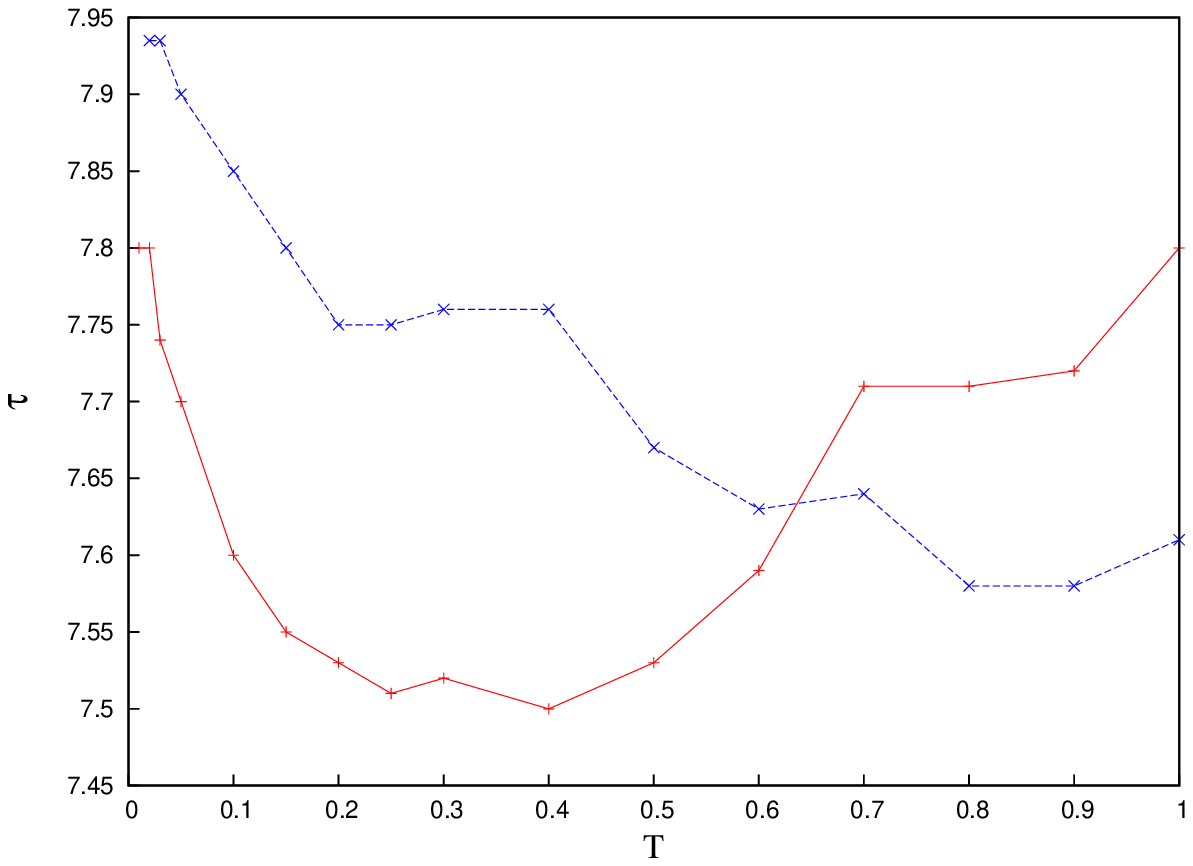}
\vspace*{-4mm}
\caption{The figure records the $\tau$ values corresponding to maximum 
$<\overline{W}>$ (in red) and  maximum $<\overline{v}>$ (in blue) as $T$ is
changed, with $F_0$ = 0.2, $\gamma_0$ = 0.07, $\lambda$ = 0.9, $\theta$ = 
0.50$\pi$.}
\end{figure}

From Fig.14, it is clear that $\overline{W}$ for LA states is much larger than 
for the SA states, one would, therefore, expect $<\overline{W}>$ to peak just 
before the SA states make their appearance around $\tau=7.5$, unless the loss 
due to appearance of SA states is more than compensated for by the rise of 
$\overline{W}$ with increase in $\tau$. However, the slope of 
$\overline{W}(\tau)$ is not so large as to offset the loss due to appearance of
the SA states, yet $<\overline{W}>$ peaks at a value of $\tau$ larger than 7.5.
Though this anomaly is curious, the peaking of $<\overline{v}>$ at low 
temperatures is more interesting. $<\overline{v}>$ peaks at a value of 
$\tau\sim 7.935$ where SA and LA states are nearly equally stable. (As 
concluded earlier, at smaller $\tau$ the LA state is more stable than SA state
and vice versa.) In other words, at low temperatures, the asymmetry in the 
probability of crossings to the left and right wells of the periodic potential 
becomes the largest at that value $\tau$ for which LA and SA states are 
equally stable!
  
Though $<\overline{W}>$ and $<\overline{v}>$ do not peak at the same $\tau$
values, the difference in $\tau$ values at which they peak is much smaller
compared to the range of $\tau$ over which the two states coexist. However,
for given values of $\gamma_0$ and $\tau$, $<\overline{W}>$ and 
$<\overline{v}>$ peak at widely separated temperatures, Fig.16. Thus the 
variation of $<\overline{W}>$ and $<\overline{v}>$ as a function of frequency 
is nearer to coincidence than the peaking of $<\overline{W}>$ (stochastic 
resonance) and $<\overline{v}>$ as a function of temperature.

As stated earlier, frictional inhomogeneity is a comparatively weak agent to 
yield ratchet effect, yet the maximum ratchet current that we obtain is not 
negligible and there is a possibility of obtaining still larger current if the 
parameters are tuned suitably. For example, we have not yet completely 
explored the variation of $<\overline{W}>$ and $<\overline{v}>$ as a function 
of $\gamma_0$ which may yield larger ratchet current.

In summary, we are closer to finding answer to all questions that we had
mentioned to begin with. We could observe stochastic resonance and ratchet 
effect in the same region of parameter space. However, SR and ratchet current 
peak do not occur  at the same temperature. We now have a better idea about the 
relative roles played by the parameters $\theta$ and $\lambda$ characterising 
the frictional inhomogeneity and also the periodicity $\tau$ of the external 
periodic field to obtain $<\overline{W}>$ and $<\overline{v}>$.
  
We thank the Computer Centre, North-Eastern Hill University, Shillong, for 
providing the high performance computing facility, SULEKOR.


\begin{thebibliography}{99}
\bibitem{Risken} H. Risken, {\underline{The Fokker-Planck Equation}} Ch. 11,
Springer-Verlag, 1989.
\bibitem{Svoboda} K. Svoboda, C.F. Schmidt, B.J. Schnapp, and S.M. Block, Nature 365, 721 (1993);
J.T. Finer, R.S. Simmons, and J.A. Spudich, Nature 368, 113 (1994).
                                                              694,1         94%
\bibitem{Reimann} P. Reimann, Phys. Rep. 361, 57 (2002).
\bibitem{Magnasco} M.O. Magnasco, Phys. Rev. Lett. 71, 1477 (1993); {\it{ibid}}
72, 2656 (1994).
\bibitem{Prost} J. Prost, J. F. Chauwin, L. Peliti, and A. Ajdari, Phys. Rev.
Lett. 72, 2652 (1994); R.D. Astumian, and M.Bier, Phys. Rev. Lett. 72, 1766
(1994).
\bibitem{Julicher} F. J\"{u}licher, A. Ajdari, and J. Prost, Rev. Mod. Phys.
69, 1269 (1996).
\bibitem{Maddox} J. Maddox, Nature 369, 181 (1994); {\it{ibid}} 369, 271
(1994).
\bibitem{Rousselet} J. Rousselet, L. Salome, A. Ajdari, and J. Prost,
Nature 370, 446 (1994).
\bibitem{Benzi} R. Benzi, A. Sutera, and A. Vulpiani J. Phys. A 14, L453
(1981).
\bibitem{Gamma} L. Gammaitoni, P. H\"{a}nggi, P. Jung, and F. Marchesoni,
Rev. Mod. Phys. 70, 223 (1998).
\bibitem{Well} T. Wellens, V. Shatokhin, and A. Buchleitner, Rep. Prog. Phys.
67, 45 (2004).
\bibitem{McN} B. McNamara, and K. Wiesenfeld, Phys. Rev. A 39, 4854 (1989).
\bibitem{Fauve} S. Fauve, and F. Heslot, Phys. Lett. A 97, 5 (1983).
\bibitem{Roy} B. McNamara, K. Wiesenfeld, and R. Roy, Phys. Rev. Lett. 60,
2626 (1988). 
\bibitem{Saikia} S. Saikia, A.M. Jayannavar, and M.C. Mahato, Phys. Rev. E 83,
061121 (2011); W.L. Reenbohn, S.S. Pohlong, and M.C. Mahato, Phys. Rev. E 85,
031144 (2012); S. Saikia, Physica A 46, 411 (2014) 
\bibitem{Wanda} W.L. Reenbohn, and M.C. Mahato, Phys. Rev. E 88, 032143 (2013).
\bibitem{Liu} K. Liu, and Y. Jin, Physica A 392, 5283 (2013).
\bibitem{Kim} Y.W. Kim, and W. Sung, Phys. Rev. E 57, R6237 (1998).
\bibitem{Buttiker} M. B\"{u}ttiker, Z. Phys. B - Condensed Matter, 68, 161
(1987).
\bibitem{Landauer} R. Landauer, J. Stat. Phys. 52, 233 (1988).
\bibitem{Blanter} Ya M. Blanter, and M. B\"{u}ttiker, Phys. Rev. Lett. 81,
4040 (1998).
\bibitem{Benjamin} R. Benjamin, and R. Kawai, Phys. Rev. E 77, 051132 (2008).
\bibitem{Statmech} W.L. Reenbohn, and M.C. Mahato, J. Stat. Mech.:Theory and
Experiment P03011 (2009).
\bibitem{overdamped} D. Dan, M.C. Mahato, and A.M. Jayannavar, Phys. Rev. E
60, 6421 (1999); D. Dan, A.M. Jayannavar, and M.C. Mahato, Int. J. Mod. Phys.
14,1585 (2000); D. Dan, M.C. Mahato, and A.M. Jayannavar, Physica A 296, 375
(2001); D. Dan, M.C. Mahato, and A.M. Jayannavar, Phys. Rev. E 63, 056307
(2001); W.L. Reenbohn, S. Saikia, R. Roy, and M.C. Mahato, Pramana - Journal of
Physics, 71, 297 (2008).
\bibitem{Qian} M. Qian, Y. Wang, and X-J. Zhang, Cin. Phys. Lett. 20, 810
(2003).
\bibitem{Wanda1} W.L. Reenbohn, and M.C. Mahato, Phys. Rev. E 91, 052151
(2015).
\bibitem{Sancho} J.M. Sancho, M. San Miguel, and D. Duerr, J. Stat. Phys. 28,
291 (1982).
\bibitem{Pramana} A.M. Jayannavar, and M.C. Mahato, Pramana - Journal of Physics,
45, 369 (1995).
\bibitem{Desloge} E.A. Desloge, Am. J. Phys. 62, 601 (1994).
\bibitem{Nume} W.H. Press, B.P. Flannery, S.A. Teukolsky, and W.T. Vetterling,
{\underline{Numerical Recipes}}, Cambridge University Press, Cambridge,
England (1987).
\bibitem{SRS} M.C. Mahato, and S.R. Shenoy, Phys. Rev. E 50, 2503 (1994);
T. Iwai, Physica A 300, 350 (2001);
M. Evstigneev, P. Reimann, C. Schmitt, and C.Bechinger, J. Phys.: Condens.
Matter 17, S3795 (2005).
\bibitem{Mannella} R.~Mannella, A Gentle Introduction to the Integration
of Stochastic Differential Equations. In : {\it Stochastic Processes in
Physics, Chemistry, and Biology}. Edited by J. A.~Freund and T. P\"{o}schel,
Lecture Notes in Physics, vol. 557, 353. Springer, Berlin, 2000.
\bibitem{Sekimoto} K. Sekimoto, J. Phys. Soc. Jpn. 66, 1234 (1997).
\end{thebibliography}
\end{document}